\documentstyle[12pt,aaspp4]{article}

\begin{document}

\title{The Cepheid Period-Luminosity relation and 
the maser distance to NGC 4258}

\author{F. Caputo \altaffilmark{1}, M. Marconi \altaffilmark{2}, 
I. Musella \altaffilmark{2}}

\affil{1. Osservatorio Astronomico di Roma, Via Frascati 33,
00040 Monte Porzio Catone, Italy; caputo@coma.mporzio.astro.it}
\affil{2. Osservatorio Astronomico di Capodimonte, Via Moiariello 16,
80131 Napoli, Italy; marcella@na.astro.it; ilaria@na.astro.it}

\begin{abstract}
In a recent paper describing HST observations of
   Cepheids in  the spiral galaxy  NGC 4258, Newman {\it et al.}
   (2001) report that  the revised calibrations and methods for the
   Key Project on the  Extragalactic Distance Scale  yield that the
   true distance modulus of this galaxy is
   $\mu_{0,4258}=29.40\pm$0.09~mag, corresponding to a metric distance
   of 7.6$\pm$0.3 Mpc. This Cepheid distance, which holds for
   18.50 mag as the true distance modulus of the  Large
   Magellanic Cloud (LMC),  is not significantly larger than
   7.2$\pm$0.5 Mpc, the value  determined by Herrnstein {\it et al.}
   (1999) from purely geometric considerations on the orbital  motions
   of water maser sources.   However, if the metallicity difference
   $\Delta$[O/H]$\sim$ 0.35 between NGC 4258 and LMC is taken into
   account,  then the Key Project methods lead to a
   metallicity-corrected value of
   $\mu_{0,4258}=29.47\pm$0.09 mag, with $\mu_{0,LMC}$=18.50 mag, 
   namely to a Cepheid   
   distance of 7.8$\pm$0.3 Mpc which is  1.2$\sigma$
   from the maser determination. One might argue that  these
   results are suggesting that a smaller distance to the LMC,
   $\mu_{0,LMC}\sim$ 18.3 mag,  should be used to calibrate the
   Cepheid brightness (see, e.g.,  Paczynski 1999). However, even though
the well known controversy between 
{\it short} and {\it long} distance scale is far from being settled and 
current determinations  of the LMC distance modulus are
 spanning $\sim$ 0.4 mag, it 
seems that the various empirical 
methods converge on  the value adopted by  the Key 
Project (see Carretta et al. 2000). In this paper we show  
   that the metallicity correction on Cepheid
   distance determinations, as   suggested by pulsation      models 
   with three
different chemical abundances  ($Y$=0.25, $Z$=0.004; $Y$=0.25, $Z$=0.008;
$Y$=0.28, $Z$=0.02), might   
   provide the natural way of reaching a close agreement   between Cepheid
   and maser distance to NGC 4258 for a wide variety of LMC distance 
determinations. 
   In fact, by adopting 
   $\mu_{0,LMC}=18.50\pm$0.10 mag we derive a     Cepheid distance of
   7.3$\pm$0.6 Mpc which is only 0.2$\sigma$ from the maser
   determination, while using $\mu_{0,LMC}=18.60\pm$0.10 mag or 
18.40$\pm$0.10 the Cepheid and maser distances are 0.8$\sigma$ 
and 0.4$\sigma$, respectively, apart. 

\end{abstract}

\keywords{stars: variables: Cepheids --
               stars:  oscillations --
               stars: distances --
	       galaxies: individual (Large Magellanic Cloud, NGC 4258)}  

%%%%%%%%%%%%%%%%%%%%%%%%%%%%%%%%%%%%%%%%%%%%%%%%%%%%%%%%%%%%%%%%%%%%%%%%%%%%%%%
\pagebreak 

\section{Introduction} 

In recent papers, we have deeply investigated the  pulsational 
properties of classical Cepheids on the basis of 
nonlinear, nonlocal and time-dependent 
convective
pulsating models where the coupling between pulsation and 
convection is properly taken into account. From computations with three 
different chemical abundances  ($Y$=0.25, $Z$=0.004; $Y$=0.25, $Z$=0.008;
$Y$=0.28, $Z$=0.02), chosen to properly cover Cepheids observed both in 
the Magellanic Clouds and in the Galaxy, the pulsation
amplitude and the average magnitudes of the pulsators are obtained  
and several predicted relations, e.g. Period-Luminosity (PL), 
Period-Color (PC),  
Period-Luminosity-Color 
(PLC) and Color-Color (CC), in the $BVRIJK$ bands are provided. 
The discussion on the input physics and computing
procedures, and the whole set of predicted relations, 
have been already presented 
(see Bono, Marconi, \& Stellingwerf  1999a; Bono {\it et al.} 1999b; Bono, Castellani, \& Marconi 2000a; 
Caputo, Marconi \& Ripepi 1999; Caputo, Marconi, \& Musella 2000a; Caputo {\it et al.} 2000b) and are not repeated. 
Here we wish only to remark that both the 
slope and zero-point of the predicted PL relations turn out to depend 
on metallicity in the sense that metal-rich pulsators 
follow  shallower PL relations and are on average fainter, 
for fixed period, than metal-poor ones. 

On this basis, the metallicity correction to a true distance 
modulus  $\mu_0$ determined from  
Cepheid PL relations at the LMC   
has been evaluated (Caputo {\it et al.} 
2000b) as given by  $\Delta\mu_0/\Delta$log$Z\sim -$0.27 
mag dex$^{-1}$, where $\Delta$log$Z$ is the difference between the 
metallicity of the Cepheids whose distance we are estimating and the 
LMC value ($Z\sim$ 0.008, see Luck {\it et al.} 
1998). This result disagrees with earlier observational clues based on 
the analysis of Cepheids in different 
fields of M31 (Freedman \& Madore 1990) and 
M101 (Kennicutt {\it et al.} 1998), which suggest an opposite 
metallicity effect on 
distance determinations (see also Kochanek 1997; Sasselov 
{\it et al.} 1997). Following Kennicutt {\it et al.} (1998), 
the metallicity 
correction is  $\Delta\mu_0/\Delta$[O/H]$\sim$+0.24
mag dex$^{-1}$, where $\Delta$[O/H] is the difference between 
the oxygen abundance of the Cepheid field and the LMC value 
([O/H]$\sim -0$.40, see Pagel {\it et al.} 1978). More recently, the 
revised procedures of the HST Key Project (Freedman {\it et al.} 2001) 
adopt a correction of 
$+0.2\pm$0.2 mag dex$^{-1}$. 

It should be noted that the empirical corrections have been 
estimated by forcing the slope of the Cepheid PL relations at different 
wavelengths to be that observed for the LMC. Since the 
pulsating models suggest that also the slope depends on the pulsator 
metallicity, we might 
suspect some systematic errors in the attempt of disentangling 
reddening from metallicity effects. 
We add that 
recent empirical studies on Galactic Cepheids seem to support 
the theoretical results. According to Groenewegen \& Oudmaijer (2000), 
the $V$ and $I$ data of 236 Cepheids from the {\it Hipparcos} catalogue 
do not exclude shallower PL relations than that observed for LMC 
Cepheids. Moreover, the $V$ and $K$ data of Cepheids in the 
Galactic open clusters with Main Sequence fitted distances 
(Hoyle, Shanks, \& Tanvir 2000) do suggest  PL$_V$ and PL$_K$ slopes of 
$\sim-$2.1 and $\sim-$2.82, respectively, significantly flatter 
than the  canonical LMC values of $-$2.76 (Freedman {\it et al.} 
2001) and $-$3.44 (Laney \& Stobie 1994). 

In any case, 
even though the LMC distance is the benchmark of 
the Cepheid-calibrated distance scale,  it would  be reasonable  
to expect that the next most important role might be 
played by the  metallicity of the observed Cepheids. 
We note that the oxygen abundance of distant galaxies with 
HST observations of Cepheids 
is on average $\sim$0.5 dex higher than that of LMC. 
Thus, ignoring the metallicity effect might lead to neglect 
corrections of $\sim -$0.1 mag (theoretical) or 
$\sim+$0.1 mag (empirical) to LMC-based distance 
moduli, with not dramatic but significant consequences on the value 
of the Hubble 
constant $H_0$.

In this context, HST observations of Cepheids in the spiral galaxy 
NGC 4258 (Maoz {\it et al.} 1999) appear of great importance since 
the distance to this galaxy has been accurately determined from direct 
measurements of the orbital motions of water maser sources 
(Herrnstein {\it et al.} 1999), providing a stringent geometrical 
test for both the HST Key Project calibration and the pulsation models. 

In the following Sect. 2 we derive
the distance to NGC 4258 relative to LMC, 
adopting the results of pulsating models, 
and we compare our {\it ``pulsational''} estimates with those 
based upon empirical Cepheid relations.
The discussion on the absolute distance to NGC 4258 and LMC   
is presented in Sect. 3, while some final comments on future work close the 
paper.

\section{Cepheid distance to NGC 4258 relative to LMC}
\subsection{Empirical determinations}

The revised Key Project method
(Freedman {\it et al.} 2001)
is based on 
the PL relations of fundamental Cepheids 

$$M_V=-1.458(\pm 0.02)-2.760(\pm 0.03)\log P\eqno(1)$$
$$M_I=-1.942(\pm 0.01)-2.962(\pm 0.02)\log P\eqno(2)$$

\noindent
where the slopes are inferred from the huge sample of LMC Cepheids in 
the OGLEII catalogue (Udalski {\it et al.} 1999), 
$P$ is the period in days, and  $M_V$ and $M_I$ are 
intensity-averaged mean 
absolute magnitudes for an assumed LMC true 
distance modulus of 18.50 mag. 

In the Key Project methodology, the 
difference in the $V$ and $I$ apparent distance moduli derived by 
fitting the above equations to the observed magnitudes of Cepheids 
gives the average value of $E(V-I)$. Adopting for the absorption in 
the $V$-band $A_V$=2.45$E(V-I)$ from the 
Cardelli {\it et al.} (1989) reddening law, 
the extinction-corrected distance 
modulus $\mu_0$ is derived. According to 
Newman {\it et al.} (2001, hereafter N2001), $V$ and $I$ 
data\footnote{According to N2001, for NGC 4258, 
we consider DoPHOT photometry  given 
the uncertainty on ALLFRAME} of Cepheids in NGC 4258 lead to  
$\mu_{0,4258}$=29.40$\pm$0.09 mag (total random error), 
namely to a distance modulus larger than the LMC value by 10.90 mag.  
  
The Key Project method is equivalent to combining 
$V$ and $I$ magnitudes to form the reddening-free  
Wesenheit quantities 
$W(VI)=I-1.45(V-I)$,  and fitting them with a linear relation 
to the period.  
Udalski {\it et al.} (1999), using a 
slightly different extinction curve which yields 
$A_V$=2.55$E(V-I)$, find that the Wesenheit-PL (hereafter WPL) 
relation of LMC Cepheids in the OGLEII catalogue is 

$$W(VI)=15.82(\pm 0.01)-3.28(\pm 0.01)\log P\eqno(3)$$
\noindent
with a standard deviation of 0.08 mag. 

Using this empirical WPL relation 
for the HST magnitudes of NGC 4258 Cepheids listed in N2001 
and adopting $\mu_{0,LMC}$=18.50 mag, 
we derive that
the Cepheid distance to this galaxy is 
$\mu_{0,4258}$=29.42 $\pm$0.09 mag (random error)$\pm$0.08 mag 
(standard deviation on WPL relation),  
in agreement with the Key 
Project result. 

Let us now consider 
that the oxygen abundance adopted by N2001 for the HST fields in NGC 4258 
is [O/H]=$-0.05\pm$0.06, which is 0.35 dex
higher than that for the LMC Cepheids. In accordance with 
Freedman {\it et al.} (2001), the  
correction to the Cepheid distance modulus, 
as based upon current observational 
results, is 0.2$\pm$0.2 mag dex$^{-1}$. This yields that the 
distance modulus we have derived for the Cepheids in 
NGC 4258 must be increased by 0.07$\pm$ 0.07 mag. 

In conclusion 
of this part of discussion, given the well known debate on the LMC distance 
(see later),  
we can state that 
the Cepheid distance 
to NGC 4258, as inferred from LMC-based PL or WPL relations
and {\it empirical} metallicity correction on the 
Cepheid distance scale, is given by 

$$\mu_{0,4258}=\mu_{0,LMC}+10.99\pm 0.14 \mbox{\ mag}\eqno(4)$$

\subsection{Theoretical models}

Concerning the results of pulsating models, let us first 
compare the above empirical PL and WPL relations with the 
predicted ones for fundamental pulsators 
with $Z$=0.008 and $Z$=0.02. 

In previous  papers
(e.g. Bono {\it et al.} 1999b; Caputo et al. 2000a), we    
have already shown that moving toward the shorter wavelengths 
the $M_{\lambda}$-log$P$ distribution of 
fundamental pulsators with mass larger than 5$M_{\odot}$, 
i.e. period longer than $\sim$ 3 days, is much better represented by
a quadratic relation ($M_{\lambda}=a+b\log P+c\log P^2$). 
However, if only periods shorter than log$P$=1.5 (as observed in 
the LMC) are considered in the 
final fit, then reliable linear solutions may be obtained. As for the 
$W(V,I)$ quantities, the predicted distributions with period are  
well represented by linear 
solutions over the whole range 0.5$\le$log$P\le$2.0.

The values of the coefficients of the theoretical $VI$ PL and WPL 
relations are given in Table 1. One finds    
that the slope of the
predicted linear PL$_V$ and PL$_I$ relations with $Z$=0.008
($-2.75\pm$0.02 and
$-2.98\pm$0.01, respectively) 
are in very good agreement with that provided by LMC Cepheids
[see Eqs.(1) and (2)]. A fair agreement is also present between the 
empirical ($-3.28\pm 0.01$) and the predicted slope ($-3.17\pm$0.04) 
of the WPL relation with  
$Z$=0.008.

Concerning the zero-points, one finds a very close agreement in the 
$I$-band [$-1.95\pm$0.01 versus $-1.94\pm$0.01 in Eq. (2)], 
whereas the predicted linear PL$_V$ relation with $Z$=0.008 
is fainter by $\sim$ 0.09 mag than Eq. (1).  
Following the Key Project procedure, this yields  
that the theoretical PL relations would give 
a larger $\mu_0$  than 18.50 mag.  
This results 
also if the Wesenheit quantities are considered: 
at log$P$=1, the  
predicted WPL relation with $Z$=0.008 gives a reddening-corrected 
magnitude of $-6.16\pm$0.11 mag, whereas from Eq. (3) one 
has $-5.96\pm$0.08 mag, 
with $\mu_{0,LMC}$=18.50 mag.

In fact, when the predicted WPL relation with $Z$=0.008 is  
applied to the LMC Cepheids provided by the OGLEII microlensing 
survey (Udalski {\it et al.} 1999), we find $\mu_{0,LMC}$=18.71$\pm$0.10 mag 
(random error)$\pm$0.10 mag (intrinsic dispersion of the theoretical 
WPL relation, Caputo {\it et al.} 2000a), 
which is 
0.2 mag larger than the Key Project adopted value.

Such a difference of 0.2 mag in the true distance modulus 
remains when the observed $V$ and $I$ data of Cepheids in NGC 4258
are taken into consideration. Based on  
the predicted WPL 
relation with $Z$=0.008, we get 
$\mu_{0,4258}=29.64\pm$0.09 mag (random error)$\pm$0.10 mag 
(intrinsic dispersion of the theoretical 
WPL relation). 

In conclusion,   
both empirical LMC-based and theoretical WPL relations  
confirm that the true distance modulus of NGC 4258 is larger than 
the LMC value  
by  10.92 mag, 
{\it on condition that the Cepheids in NGC 4258 have the same 
metallicity as for the LMC.}

However, the pulsating models suggest  
that for fixed period the 
metal-rich  
pulsators are generally fainter than metal-poor ones,  with 
the difference in luminosity increasing towards the longer periods. 
As a consequence, if 
a higher metal content is adopted, 
then the true distance modulus inferred
from the predicted relations 
should decrease, with the effect of metallicity increasing as 
the average period of the Cepheid sample moves towards longer periods. 
In fact, for the typical value of Galactic Cepheids, 
i.e. $Z$=0.02, we determine 
$\mu_{0,LMC}=18.65\pm$0.14 mag (total error) and 
$\mu_{0,4258}=29.51\pm$0.14 mag.
 
Eventually, in accordance with
the current oxygen abundance estimates for LMC ([O/H]$\sim-$0.4) 
and NGC 4258
([O/H]=$-0.05\pm$0.06), adopting 
$\Delta$log$Z$=$\Delta$[O/H] and    
interpolating between the $Z$=0.008 and $Z$=0.02  
results for NGC 4258, 
we derive that the predicted  
{\it metallicity-corrected}  Cepheid 
distance modulus of NGC 4258  
is given by 

$$\mu_{0,4258}=\mu_{0,LMC}+10.82\pm 0.14 \mbox{\ mag},\eqno(5)$$

\noindent
which is 0.17 mag lower than the value determined with the empirical 
metallicity correction [see Eq. (4)],  
{\it for any adopted LMC distance.}

\section{Absolute distances}

According to Herrnstein {\it et al.} (1999),
very-long-baseline-interferometry observations of the NGC 4258 maser
provide the way to determine a metric distance of 7.2$\pm$0.5 Mpc, as
inferred from simple geometric considerations of the speed and motions
of the maser. This number, corresponding to a true distance modulus
$\mu_{0,4258}$=29.28$\pm$0.15 mag, appears the most accurate
distance to faraway galaxies  so far measured.

As for the Cepheid distance, its determination via Eq. (4) and Eq. (5) 
requires the knowledge of the absolute distance to the LMC. 
Given the well known controversy between  {\it short} and  
{\it long} distance scale, 
let us refer to the excellent reviews by Walker (1999) and 
Carretta et al. (2000) which provide a synthesis of LMC distance 
determinations. 

As a whole (see also Fig. 1 in Clementini et al. 2000), the various distance 
indicators suggest a distance modulus in the range of
$\sim$ 18.3 mag to $\sim$ 18.7 mag and, according to the authors, they  
seem to converge on   
18.54$\pm$0.07 mag (Carretta et al. 2000) or 18.50$\pm$0.07 (Clementini et 
al. 2000).

In the upper panel of Fig. 1 we 
show how any change in the LMC distance will affect the determination 
of the Cepheid distance to NGC 4258. 
By inserting  $\mu_{0,LMC}=18.50\pm$0.10 mag into 
Eq. (4) and Eq. (5),  
we derive a     Cepheid distance of
7.9$\pm$0.6 Mpc (empirical metallicity correction) and 
   7.3$\pm$0.6 Mpc (theoretical metallicity correction), which are 
1.2 $\sigma$ and   
0.2$\sigma$, respectively,  from the maser determination. 

Since asserting 
the true distance to the LMC is a tricky matter to handle, 
also in consideration 
of the consequent calibration of secondary distance indicators and 
measurement of the Hubble constant, let us turn the problem around, so 
that the maser distance to NGC 4258 is used to recalibrate the 
Cepheid brightness. As shown in the lower panel of Fig. 1, the 
agreement within 1$\sigma$ 
between the Cepheid and the maser distance to 
NGC 4258 
requires $\mu_{0,LMC}$ in the range of 18.15$-$18.45 mag or 18.35$-$18.60, 
depending on whether the empirical or theoretical metallicity correction 
is adopted.

\section{Final comments}

In the previous section we have shown that, 
if the maser distance to NGC4258 is 
adopted to recalibrate the Cepheid PL relations, then 
the predictions of pulsation models appear more
suitable,  in comparison with the
results of empirical PL relations and metallicity correction, 
to account for our current knowledge of the
LMC distance, supporting a distance modulus close to 
18.50 mag.

However, we cannot fail to mention that such a value appears   
lower than 
the  
number (18.7$\pm$0.1 mag) suggested both by {\it Hipparcos} parallaxes of 
Galactic Cepheids (see Feast 1999) and the pulsating models 
with $Z$=0.008 (see Sect. 2). 

On the observational side, we can  note that
the {\it Hipparcos} parallaxes
could not fix the star magnitudes to the few percent accuracy
that is desirable.  Moreover, Groenewegen \& Oudmaijer (2000) find that
the PL relation in the $K-$band gives  a shorter LMC distance than $VI$
relations, and the difference can be as large as $\sim$ 0.2 mag.

On the theoretical side, even though our pulsating 
models adopt the most updated physics, there are still several 
systematic uncertainties due to the adopted ratio $\Delta Y/\Delta Z$, 
adopted mass-luminosity relation, bolometric corrections and 
temperature-color transformations. 

The chemical compositions of the models were fixed following a 
primordial helium content $Y_p$=0.23 and  
$\Delta Y/\Delta Z\sim$ 2.5. In 
order to scrutinize the effect of helium, new computations 
have been performed adopting a higher ratio ($\sim$ 4). Preliminary results 
with $Y$=0.31 and $Z$=0.02
(Fiorentino 2001, degree thesis)  show that 
an increased $\Delta Y/\Delta Z$ might reduce the metallicity effect on Cepheid PL relations.

The adopted mass-luminosity relation 
for pulsation models is based on the canonical 
(no overshooting) evolutionary computations 
by Castellani, Chieffi, \& Straniero (1992). Taking into account the most 
updated canonical mass-luminosity relation provided by 
Bono {\it et al.}  (2000b), the expected bolometric magnitude differences, 
at fixed mass, 
range from 0.08 to 0.25 
mag depending on the stellar mass. However, as discussed in 
Bono {\it et al.}  
(1999b), the mild 
overshooting models, which are brighter by 0.62 mag than the canonical ones, at fixed mass,
would give rather similar PL relations, at least
with log$P\le$ 1.8. As a whole, 
the adopted mass/luminosity relations seem to play a
minor role in determining the pulsational properties of Cepheids.

Concerning model atmospheres, the results presented in this paper 
rely on bolometric corrections and colors by Castelli, Gratton, \& Kurucz    
(1997a,b). The comparison of the adopted $V-I$ colors 
for solar metallicity with the more updated evaluations by 
Bessell, Castelli, \& Plez (1998a,b) shows that differences are 
negligible. 

The remaining source of uncertainty in pulsation models 
deals with the treatment of convection and in particular 
the mixing length parameter ($\alpha$) adopted to close the non 
linear equation system (see Bono \& Stellingwerf 1994 for details). 
The topology of the instability strip is affected by the $\alpha$  
parameter in the sense that the strip narrows as $\alpha$ increases, 
with a larger effect on the red boundary (up to 300 K for an 
increase of 0.3 in the $\alpha$ parameter). Since the PL relations 
sensitively depend on the edges of the pulsation region (i.e. 
the distribution of pulsators 
within the instability strip), variations of the $\alpha$ parameter 
could in principle affect the PL relation coefficients, but with minor 
effects on the WPL relation (see Fig. 12 in Caputo {\it et al.} 
2000a). 
However, the dependence of convective efficiency on metal abundance 
might alter the metallicity effect on the Cepheid 
pulsation properties. Important clues on this critical 
issue are expected from the comparison between observed and 
modeled morphology of light and radial velocity curves 
(Bono {\it et al.}   in preparation; see also
Bono, Castellani, \& Marconi  2000c).

In conclusion, we cannot exclude that future refinements in 
pulsating models and more 
accurate parallaxes of Galactic Cepheids might lead to 
a LMC true distance modulus  close to 18.50$\pm$0.10 mag, 
the value which is suggested by empirical non-Cepheid methods.

One agrees that all the above discussion is based upon the assumption 
that the maser distance 
to NGC 4258 is 
correct. 
As presently we can rely on this unique maser distance, 
we hope that in the next future 
such a promising technique would 
increase the number of distance determinations to external galaxies
with Cepheid observations. This seems fundamental for testing 
the whole pulsational scenario, as well as for setting 
the distance to LMC. 

As for NGC 4258 itself,  
the number of observed Cepheids is 
rather small (15), so that new observations are needed to 
improve the statistics and reduce the uncertainties on 
PL distance determinations.

\acknowledgments

We want to thank our referee who helped us to improve the paper. 
Financial support for this work was provided by MURST under the scientific
 project ``Stellar observables of
cosmological relevance''.

%%%%%%%%%%%%%%%%%%%%%%%%%%%%%%%%%%%%%%%%%%%%%%%%%%%%%%%%%%%%%%%%%%%%%%%%
\pagebreak

%%%%%%%%%%%%%%%%%%%%%%%%%%%%%%%%%%%%%%%%%%%%%%%%%%%%%%%%%%%%%%%%%%%%%%%%
%                         Table(s)  
%%%%%%%%%%%%%%%%%%%%%%%%%%%%%%%%%%%%%%%%%%%%%%%%%%%%%%%%%%%%%%%%%%%%%%%%

\pagebreak

\begin{table}

\begin{center}

\begin{tabular}{cccc}
\multicolumn{4}{c}{TABLE 1.}\\
\multicolumn{4}{c}{Theoretical 
relations for fundamental pulsators}\\
\hline
\hline
Z & $a$ & $b$ & $c$  \\
\hline
&&&\\
\multicolumn{4}{c}{$M_V=a+b\log P+c\log P^2$}\\
&&&\\
0.008 & -0.86 $\pm$ 0.04 & -3.98 $\pm$ 0.09 & +0.67 $\pm$ 0.05  \\
0.02  & -1.41 $\pm$ 0.03 & -2.75 $\pm$ 0.07 & +0.30 $\pm$ 0.03  \\
\hline
&&&\\
\multicolumn{4}{c}{$M_I=a+b\log P+c\log P^2$}\\
&&&\\
0.008 & -1.59 $\pm$ 0.03 & -3.84 $\pm$ 0.07 & +0.47 $\pm$ 0.03  \\
0.02  & -1.98 $\pm$ 0.02 & -2.99 $\pm$ 0.05 & +0.23 $\pm$ 0.02  \\
\hline
&&&\\
\multicolumn{4}{c}{$M_V=a+b\log P$}\\
&&&\\
0.008 & -1.37 $\pm$ 0.03 & -2.75 $\pm$ 0.02 & - - \\
0.02  & -1.62 $\pm$ 0.01 & -2.22 $\pm$ 0.01 & - - \\
\hline
&&&\\
\multicolumn{4}{c}{$M_I=a+b\log P$}\\
&&&\\
0.008 & -1.95 $\pm$ 0.01 & -2.98 $\pm$ 0.01 & - -\\
0.02  & -2.14 $\pm$ 0.01 & -2.58 $\pm$ 0.01 & - - \\
\hline
&&&\\
\multicolumn{4}{c}{$W(V,I)=a+b\log P$}\\
&&&\\
0.008 & -2.99 $\pm$ 0.10 & -3.17 $\pm$ 0.04 & - - \\
0.02  & -3.04 $\pm$ 0.10 & -3.02 $\pm$ 0.04 & - - \\

\hline
\end{tabular}

\end{center}
\begin{minipage}{1.00\linewidth}

Note. --- Theoretical PL$_V$, PL$_I$ and $WPL$ relations 
as a function of the metal content $Z$. The linear solutions 
of the PL relations  
refer to pulsators with log$P<$1.5
\end{minipage}

\end{table}

%%%%%%%%%%%%%%%%%%%%%%%%%%%%%%%%%%%%%%%%%%%%%%%%%%%%%%%%%%%%%%%%%%%%%%%%
%                           Figure Captions
%%%%%%%%%%%%%%%%%%%%%%%%%%%%%%%%%%%%%%%%%%%%%%%%%%%%%%%%%%%%%%%%%%%%%%%%
\pagebreak

%fig.1%
%\psfig{figure=zp.ps,width=8truecm}
\figcaption{(upper panel) The Cepheid distance modulus of NGC 4258 as a function of the 
distance to the LMC. Dashed and dotted lines define the distance range, as 
predicted by empirical 
[Eq. (4)] and 
theoretical metallicity correction [Eq. (5)], respectively.
The solid lines show the limits of the maser 
($\mu_{0,4258}=29.28\pm$0.15 mag) distance. (lower panel) The difference 
between Cepheid and maser distance to NGC 4258 (in $\sigma$ units) as a function of 
the distance to the LMC.  The solid lines refer to a 1$\sigma$ difference.}


\begin{references}
\reference{} Bessell, M. S., Castelli, F., \& Plez, B. 1998, A\&A, 333, 231
\reference{} Bessell, M. S., Castelli, F., \& Plez, B. 1998, A\&A, 337, 321
\reference{} Bono G., Caputo F., Castellani V., \& Marconi M. 1999b, ApJ, 512, 711
\reference{} Bono, G., Caputo, F., Cassisi, S., Marconi, M.,
 Piersanti, L., \& Tornamb\`e, A. 2000b, ApJ, 543, 955
\reference{} Bono G., Castellani V., \& Marconi M. 2000a, ApJ, 529, 293  
\reference{} Bono G., Castellani V., \& Marconi M. 2000c, ApJ, 532, L129  
\reference{} Bono G., Marconi M., \& Stellingwerf R.F. 1999a, ApJS, 122, 167
\reference{} Bono, G. \& Stellingwerf, R. F. 1994 ApJS, 93, 233
\reference{} Caputo F., Marconi M., \& Musella, I. 2000a, A\&A, 354, 610 
\reference{} Caputo F., Marconi M., Musella I., \& Santolamazza P.  2000b, 
A\&A, 359, 1059
\reference{} Caputo F., Marconi M., \& Ripepi V. 1999, ApJ, 525, 784
\reference{} Cardelli, J.A., Clayton, G.C., Mathis, J.S. 1989, ApJ, 345, 245
\reference{} Carretta, E., Gratton, R.G., Clementini, G., Fusi Pecci, F. 2000, ApJ, 533, 215
\reference{} Castellani,  V., Chieffi, A. \& Straniero, O. 1992, ApJS, 78, 517
\reference{} Castelli, F., Gratton, R. G., \& Kurucz, R. L. 1997a, A\&A, 318, 841
\reference{} Castelli, F., Gratton, R. G., \& Kurucz, R. L. 1997b, A\&A, 324, 432
\reference{} Clementini, G., Gratton, R.G., Bragaglia, A., Carretta, E., Di Fabrizio, L., 2000, astro-ph/0007471
\reference{} Feast M.W. 1999, PASP, 111,775
\reference{} Freedman, W. L. et al. 2001, ApJ, 553, 47
\reference{} Freedman, W. L. \& Madore, B. F. 1990, ApJ, 365, 186
\reference{} Groenewegen M.A.T. \& Oudmaijer R.D. 2000, A\&A, 356, 849
\reference{} Herrnstein, J. R. et al. 1999, Nature, 400, 539
\reference{} Hoyle, F., Shanks, T., \& Tanvir, N. R. 2000, MNRAS, submitted (astro-ph/0002521)
\reference{} Kennicutt, R. C., et al. 1998, ApJ, 498, 181
\reference{} Kochanek, C. S. 1997, ApJ, 491, 13
\reference{} Laney, C. D. \& Stobie, R. S. 1994, MNRAS, 266, 441
\reference{} Luck, R. E., Moffett, T. J., Barnes, T. G.,\& Gieren, W. P. 1998, ApJ, 115, 605
\reference{} Maoz, E., Newman, J. A., Ferrarese, L., Stetson, P. B., Zepf, S. E., Davis, M., Freedman, W. L., \& Madore, B. F. 1999, Nature, 401, 351
\reference{} Newman, J.A., Ferrarese, L., Stetson, P.B., Maoz, E.,  Zepf, S.E., Davis, M., Freedman, W.L.,  \&  Madore B.F.  2001, ApJ,  accepted (astro-ph/0012377)[N2001]
\reference{} Paczynski, B. 1999, Nature, 401, 331
\reference{} Pagel, B. E. J., Edmunds, M. G., Fosbury, R. A. E.,
\& Webster, B. L. 1978, MNRAS, 184, 569
\reference{} Sasselov, et al. 1997, A\&A, 324, 471
\reference{} Udalski, A., Soszynski, I., Szymanski, M., Kubiak, M.,
 Pietrzynski, G., Wozniak, P., \& Zebrun, K. 1999, AcA, 49, 223
\reference{} Walker A.,  1999, in Post-Hipparcos Cosmic Candles, eds.
A. Heck, F. Caputo (Dordrecht, Kluwer Academic Publishers), p. 125
\end{references}
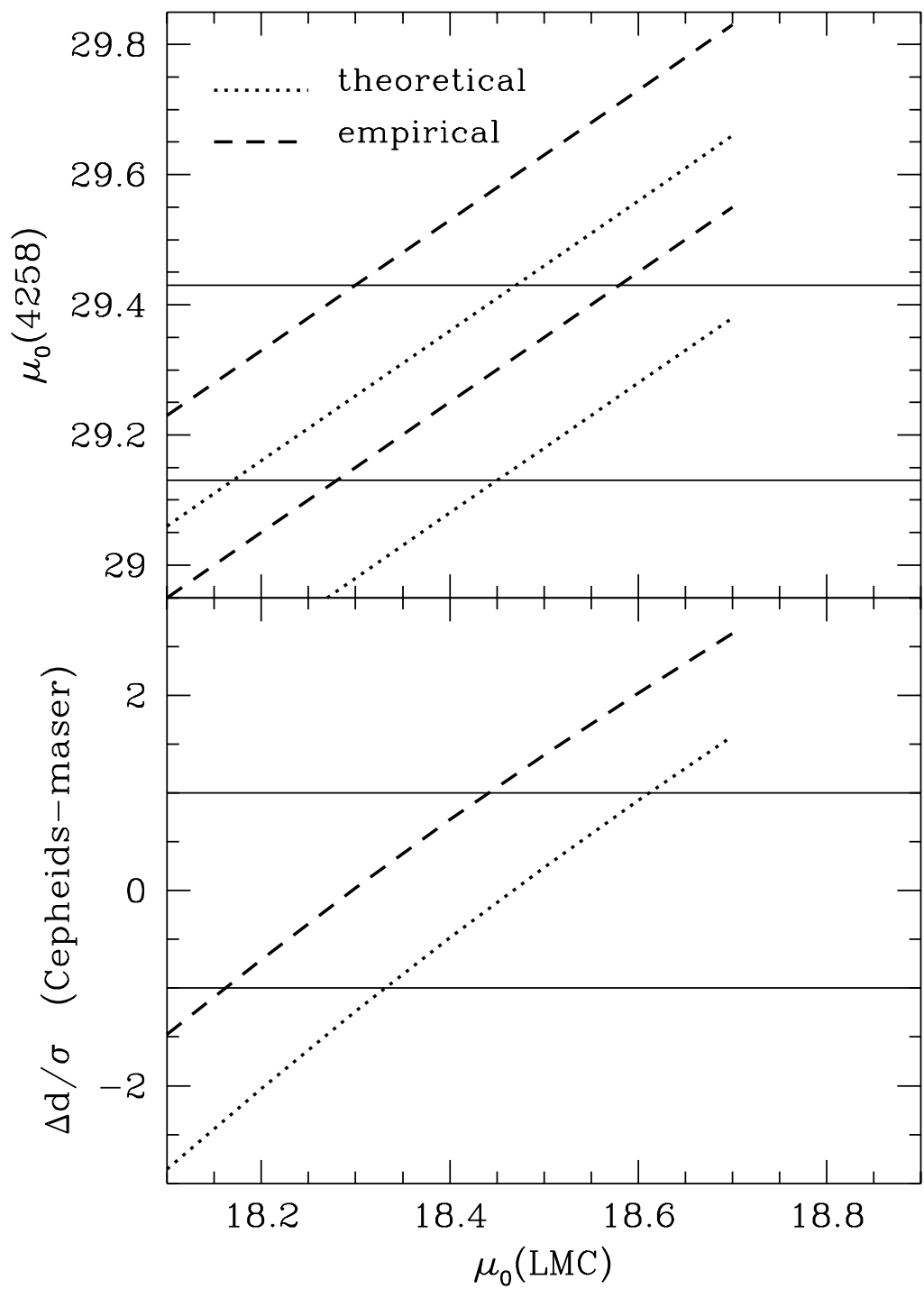
\end{document}